\documentclass[sigconf]{acmart}

\usepackage{xspace}
\usepackage{enumitem}

\usepackage{todonotes}

\usepackage[utf8]{inputenc}

\usepackage{booktabs}
\usepackage{tabularx}
\usepackage{array}
\usepackage{makecell}

\usepackage{graphicx}
\usepackage{subcaption}

\AtBeginDocument{%
  }

\copyrightyear{2027}
\acmYear{2027}
\setcopyright{cc}
\setcctype{by}
\acmConference[SIGCSE TS 2027]{Proceedings of the 58th ACM Technical Symposium on Computer Science Education V.1}{February 17--20, 2027}{Sacramento, CA, USA}
\acmBooktitle{Proceedings of the 58th ACM Technical Symposium on Computer Science Education V.1 (SIGCSE TS 2027), February 17--20, 2027, Sacramento, CA, USA}
\acmDOI{10.1145/3857222.3860154}
\acmISBN{979-8-4007-3075-7/2027/02}

\newcommand{\compassx}{CompassX\xspace}
\newcommand{\tpas}{$TreatPAs$\xspace}
\newcommand{\cpas}{$ControlPAs$\xspace}

\newcommand{\signedtime}{ $\Delta time$\xspace}
\newcommand{\signeddiffi}{ $\Delta difficulty$\xspace}

\begin{document}




\title[Better Plans, Better Outcomes: Integrating Historical Data into Student Planning in a Computing Course]{Better Plans, Better Outcomes: Integrating Historical Data into Student Planning in a Computing Course}






\author{Jiaen Yu}
\affiliation{%
  \institution{University of California, San Diego}
  \city{La Jolla}
  \country{United States}
}
\email{jiy037@ucsd.edu}

\author{Jinya Jiang}
\affiliation{%
  \institution{University of California, San Diego}
  \city{La Jolla}
  \country{United States}
}
\email{j9jiang@ucsd.edu}

\author{Soohyun Nam Liao}
\affiliation{%
  \institution{University of California, San Diego}
  \city{La Jolla}
  \country{United States}
}
\email{snam@ucsd.edu}

\renewcommand{\shortauthors}{Jiaen Yu, Jinya Jiang, \& Soohyun Nam Liao}

\begin{abstract}

Effective planning is critical for academic success, but in many computing courses students often lack accurate benchmarks for assessing the difficulty and time required for academic tasks. This phenomenon is more pronounced in large courses where instructors have limited capacity to provide individualized guidance on planning and task estimation. Although many planning tools exist, still, they rarely leverage task-specific, empirically grounded data from actual course contexts. Thus, we explored whether providing task-specific, data-driven information---assignment previews and cohort-level historical statistics---could support more effective planning. We present a lightweight and data-driven planning intervention embedded in the programming assignment workflow of an undergraduate computing course. We conducted quantitative and qualitative studies with 252 students across five programming assignments. Our results indicated that, in authentic course settings, the intervention led to more conservative and executable plans, with significantly higher plan adherence. Students with stronger adherence also achieved better performance on both programming assignments and exams. Qualitative findings also showed both practical planning support and mental benefits from intervention. These results highlight the potential of data-driven intervention to support student planning in larger courses settings where individual guidance from an instructor is further limited.

\end{abstract}







\maketitle

\section{Introduction}

Effective planning is fundamental to academic success~\cite{pintrich2000role, tan2008correlates}. When students fail to plan effectively, they may develop inefficient learning strategies, experience increased procrastination, and achieve suboptimal learning outcomes~\cite{hartwig2012study, goda2015procrastination, tan2008correlates}. 
Planning involves multiple components, among which task estimation---predicting the time, effort, and difficulty required to complete a task---is particularly important~\cite{A_Cognitive_Model_of_Planning, miller2017plans}. 
However, prior work has shown that students consistently struggle to accurately estimate the time and difficulty required for academic tasks~\cite{buehler1994exploring}.
This challenge is especially pronounced in large computing courses, where instructional team's guidelines are often generic and personalized feedback on programming assignments is difficult to provide in a timely manner.

Various planning support tools exist, from generic calendar applications to sophisticated LLM-powered systems. 
However, most suffer from a similar limitation in educational contexts: they lack task-specific, empirically grounded data about actual academic task requirements.
This limitation leaves students to rely on uninformed guesses or generic heuristics when estimating workload and difficulty of assignment, potentially contributing to procrastination, stress, and poorer learning outcomes~\cite{hartwig2012study, goda2015procrastination}.

To address this gap, we present a lightweight planning intervention that integrates easy-to-collect data directly into students' programming assignment planning workflow. 
The intervention presents assignment previews together with aggregate cohort statistics from previous course offerings, including time spent distributions and difficulty ratings. By providing task-specific benchmarks before students begin an assignment, the intervention aims to support more realistic planning and workload estimation.
Our research questions are as follows:

\begin{itemize}
    \item RQ1: How does providing historical data impact students’ planning adherence?
    \item RQ2: How does planning adherence relate to students’ learning outcomes?
    \item RQ3: How do students perceive this data-driven intervention?
\end{itemize}

The intervention was implemented on an existing web application named \compassx, and was deployed in a CS2 course (Section~\ref{sec:3_intervention_design}). 
It displayed programming assignment previews together with historical cohort statistics right when students set up their plans, allowing them to use task-specific benchmarks during workload estimation and scheduling.






We investigated the effectiveness in planning of our intervention and how it
relates to learning outcomes in an undergraduate data structures course (Section~\ref{sec:4_prototype_evaluation}) with 252 students. 
Our findings demonstrate that this data-driven support for task estimation leads to more conservative and achievable plans. 
Students who highly adhered to their plans also achieved better learning outcomes than those who did not (Section~\ref{sec:result_behavior_performance}).  
The responses from the end-of-term student survey also indicated that more students rated their experience as positive than negative. 
They even highlighted mental health benefits as well as practical planning benefits (Section~\ref{sec:6_result_of_thamatic_analyses}). 
Despite such positive results, our results also revealed some challenges with our design. This paper describes those challenges in detail and how they inform future iterations as well.
Still, our findings highlight the potential of historical cohort data to serve as a scalable approach for supporting planning in computing education.

\section{Related Work}









\subsection{Planning in Learning Contexts}
\label{sec: effective_planning_process}


Planning is a key component of Metacognition~\cite{flavell1979metacognition, schraw1995metacognitive}, which has long been identified as a key predictor of student success~\cite{loksa2022metacognition}.
In learning contexts, planning plays a critical role in how students allocate time, effort, and strategies to achieve academic goals~\cite{hartwig2012study, goda2015procrastination, tan2008correlates}.
According to foundational work, effective planning requires setting clear and specific goals, accurately assessing available resources, and making realistic estimates of time, effort, and strategies needed to complete tasks \cite{A_Cognitive_Model_of_Planning, miller2017plans}. 
Building on this foundation, researchers further elaborated that effective planning specifically involves "selecting appropriate strategies and allocating resources that affect performance" \cite{schraw1995metacognitive}.
This perspective highlights that effective planning is not just about creating schedules, but about developing self-awareness of one's learning processes and capabilities.

Despite the importance of effective planning, students often struggle to accurately estimate the time and effort required to complete academic tasks~\cite{steel2007nature, buehler1994exploring}.
Such misestimation can contribute to procrastination, increased stress, and poorer academic outcomes \cite{hartwig2012study, goda2015procrastination, tan2008correlates}.
Research on the "planning fallacy" further shows that individuals consistently underestimate task completion times, even when they have prior experience with similar activities \cite{buehler1994exploring}.

These challenges are amplified in large-enrollment courses, where opportunities for individualized guidance and feedback are limited \cite{gibbs2014teaching, mulryan2010teaching, nicol2006formative}. As a result, students are often expected to self-regulate their learning while relying on largely generic planning support, motivating the need for more task-specific planning interventions with personalized guidance for their planning efforts.

\subsection{Planning Support Interventions}
\label{sec:planning_support_tools}

Planning support interventions have evolved significantly from conventional approaches that provide general guidelines and awareness tools to more recent LLM-powered systems that offer personalized guidance. 
The existing interventions generally fall into three categories: guideline-oriented approaches, awareness-oriented approaches, and LLM-powered approaches.

\textbf{\emph{Guideline-oriented approaches}}
introduce learners to general principles of productivity, organization, and/or time management, such as calendars and to-do lists \cite{palen1999social, joe2002augmenting, kincaid1985electronic}.
For example, TaskVista \cite{bellotti2004studies} and ScriPer \cite{haraty2012individual} followed user-centered principles to accommodate individual differences in task management. Lund and Wiese's planning application follows Time Management Planning (TMP) principles to promote systematic planning \cite{lund2021less}. 
While these approaches offer broadly applicable planning guidance, they remain largely generic \cite{bellotti2004studies, haraty2012individual, lund2021less, palen1999social, joe2002augmenting} and provide limited support for estimating the demands of specific academic tasks.

\textbf{\emph{Awareness-oriented approaches}} 
help learners reflect on their behaviors and explore solutions through progress tracking and visualization. 
For example, Proccoli~\cite{zhao2023curb} and ClearMind~\cite{li2025exploring} support reflection on procrastination behaviors, as well as Group Spinner \cite{kharrufa2017group} and Metafora Analytics \cite{harrer2013analytics}, which group participation and progress to promote group accountability.
\compassx \cite{yu2025student, yu2025engagement} similarly supports planning and reflection by prompting students to create plans before assignments and evaluate plan execution afterward. 
While these approaches encourage reflection, they generally do not provide suggestions for future planning and often rely only on data generated by the current user.


\textbf{\emph{LLM-powered approaches}} 
have been increasingly explored to support learning and educational tasks~\cite{neumann2024llm, yu2026improving}, including plan generation, task decomposition, and refinement~\cite{he2025plan, huang2022inner, yao2023react, raman2024cape}.
Systems such as CAPE \cite{raman2024cape}, ReAct \cite{yao2023react} 
illustrate the potential of LLMs to offer flexible and individualized planning support at scale~\cite{rivera2024iterative}. 
However, their planning abilities are mostly based on common-sense tasks rather than course- or task-specific contexts \cite{valmeekam2023planbench}, making it difficult to determine whether generated plans are realistic and executable for particular academic tasks. 
Furthermore, LLM-generated plans may contain hallucinations \cite{ji2023survey}, compromising their reliability \cite{cohn2024believing, chen2025evaluating}. 
Crucially, existing LLM-based approaches rarely leverage the rich empirical data---the experiences of previous students who completed the same assignments in the same course context.

None of the above mentioned work leverages what may be the most valuable as realistic benchmarks: historical data from previous cohorts who have completed identical or very similar tasks. 
In this paper, we propose an intervention that surfaces aggregated historical data within the programming assignment planning workflow to provide students with more realistic benchmarks for planning. 

\section{Intervention Design}
\label{sec:3_intervention_design}

\begin{figure*}
  \centering
  \includegraphics[width=0.9\textwidth]{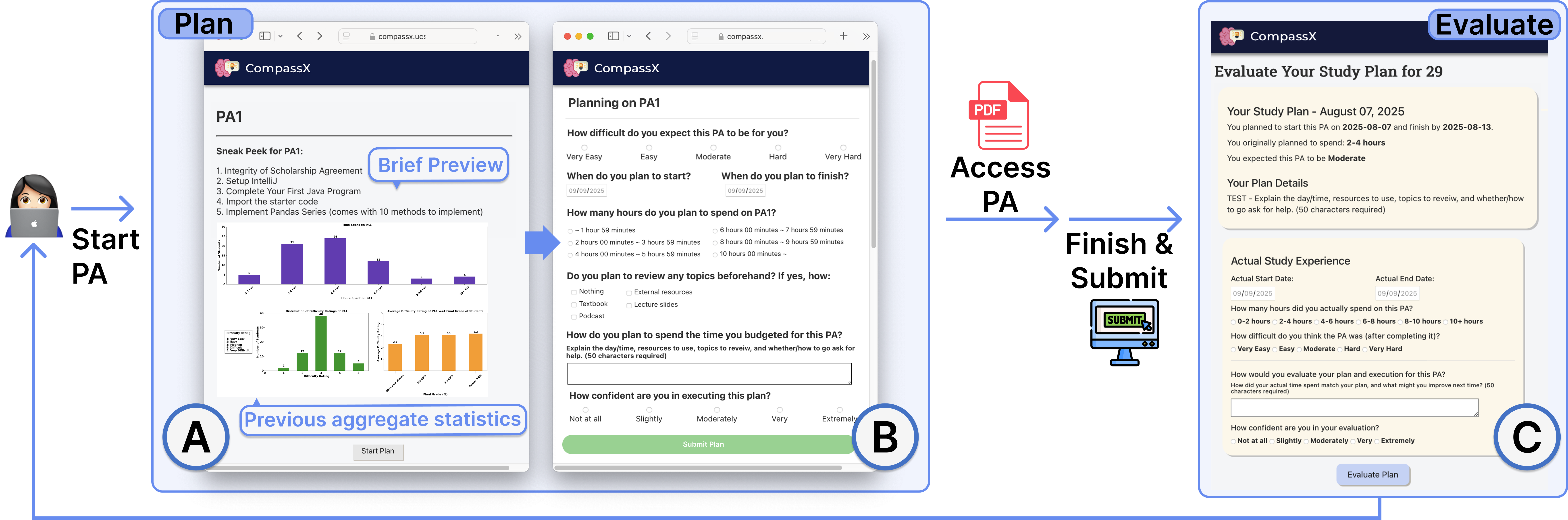}
  \caption{Workflow of the Data-driven Intervention: (A) PA preview page with preview and aggregate statistics from previous cohorts. (B) PA planning page with structured prompts for difficulty estimation, time planning, and start/end dates. (C) PA evaluate page displaying students' original plans alongside actual outcomes for reflection and comparison.}
  \Description{Data-driven intervention showing historical PA statistics during the planning phase. (A) PA preview page with aggregate statistics from previous cohorts including difficulty and time distributions. (B) PA planning page with structured prompts for difficulty estimation, time planning, start/end dates, and confidence assessment. (C) PA evaluate page displaying students' original plans alongside actual outcomes for reflection and comparison.}
  \label{fig:intervention}
\end{figure*}

We integrated a data-driven intervention into an undergraduate CS2 course to support students' planning on their Programming Assignments (PA).  
The intervention utilized two types of information: 1) aggregate cohort statistics and 2) a brief preview of an assignment. 
The statistics featured overall distribution of time spent and difficulty ratings from a past course offering in Winter 2025 (N = 69), plus average difficulty ratings with respect to different final grades. 
The preview provided the structure of a PA and a list of key tasks to complete. Figure~\ref{fig:intervention}(a) presents an example of the intervention.

\begin{figure}
  \centering
  \includegraphics[width=0.46\textwidth]{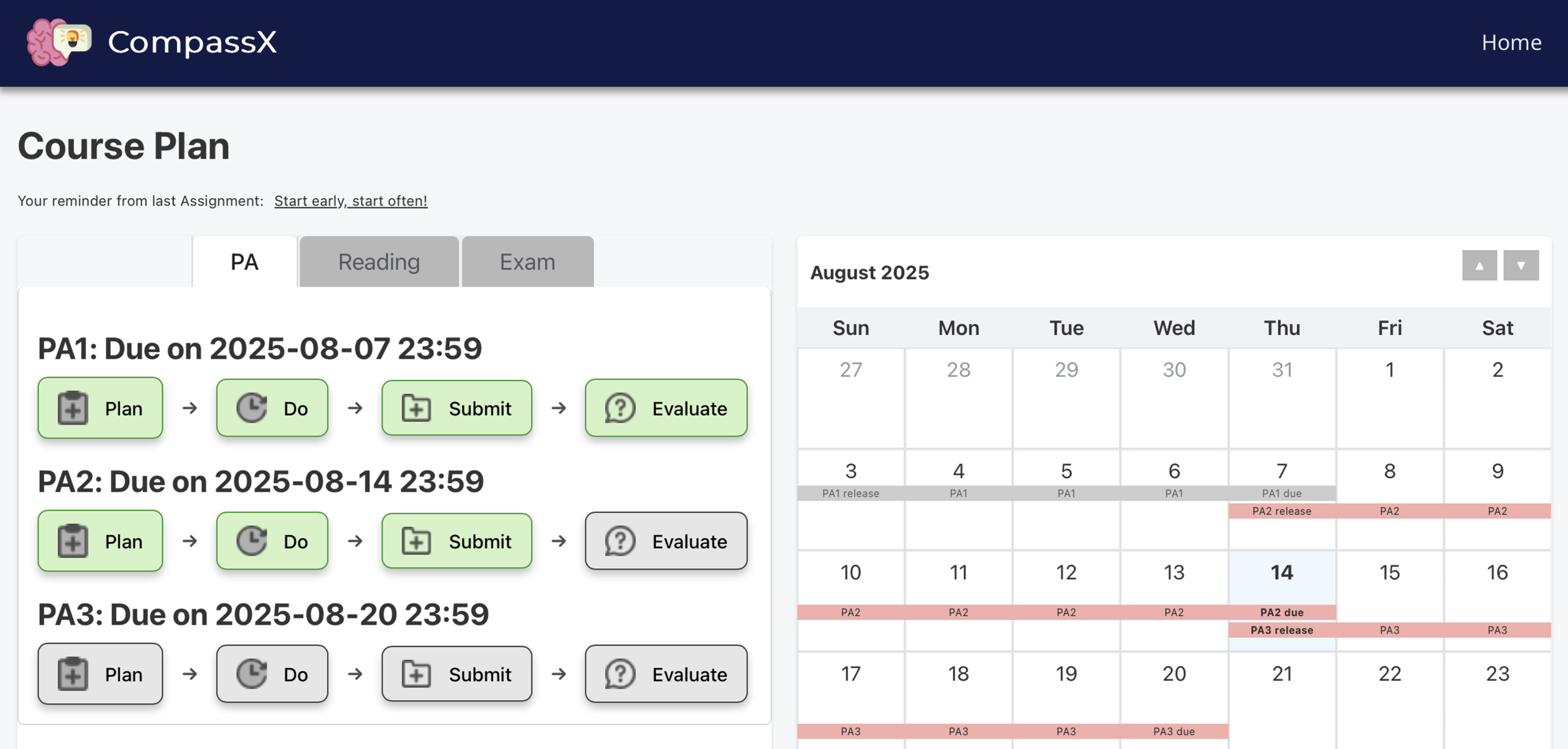}
  \caption{CompassX interface with a sequential workflow.}
  \Description{CompassX platform interface displaying the sequential Plan-Do-Submit-Evaluate workflow for programming assignments on the left side, with calendar integration showing assignment timelines and due dates on the right side.}
  \label{fig:interface}
\end{figure}



The intervention was deployed through \compassx (Figure \ref{fig:interface}), a web-based learning platform that already offers planning and evaluation support features as part of the assignment workflow (Section~\ref{sec:planning_support_tools}). 
We modified the platform to implement the intervention within the existing workflow.

Figure \ref{fig:intervention} illustrates the three-stage intervention workflow that students experience when accessing PAs through the enhanced \compassx. 
The workflow begins with a preview page (Stage A) presents the preview of a PA and historical cohort statistics, including distributions of difficulty ratings and time spent.
An entire screen is dedicated to data presentation, making engagement with data-driven benchmarks a required step in the planning process.

Students then proceed to a planning interface (Stage B), where they estimate assignment difficulty and expected time, select planned start and end dates, and identify any necessary preparation work.
Historical data was incorporated to help students calibrate their estimation, while an open-ended question prompts students to consider their current workload, course commitments, and scheduling constraints.

Only after students submit their detailed plans can they access the actual assignment and upload their completed work. Upon assignment completion, students encounter the evaluation interface (Stage C), where their original plans and estimates are displayed alongside actual outcomes to support reflection on plan execution.



This workflow replaced immediate assignment access with a structured sequence that prioritizes task estimation using realistic benchmarks from prior cohorts. 
By embedding the intervention into \compassx, planning support becomes part of the existing assignment workflow. 
At the same time, the platform captures both students’ planning intentions and actual outcomes, enabling continuous refinement of historical benchmarks for future students.


\section{Intervention Evaluation Methods}
\label{sec:4_prototype_evaluation}


\subsection{Evaluation Setup}

Our intervention was tested in an undergraduate-level data structures course (CS2) offered at a public R1 university during Spring 2025. 
The course included five PAs, reading quizzes, two midterms, and one final project. 
Analyses focused on programming assignments, where the intervention was applied, and Midterm 1 as a learning-outcome measure. Quizzes and the final project were excluded because the intervention was not applied to them, while Midterm 2 was excluded due to an academic integrity incident.

We used a within-subject design in which all students worked on the identical set of PAs; and three PAs included our intervention with cohort-based statistics (i.e. treatment PAs, denoted as \tpas) while two PAs did not (i.e. control PAs, denoted as \cpas), reducing inter-individual variability.
We alternated \cpas and \tpas throughout the term in order to control for potential ordering effects. Specifically, PA1, 3, and 5 are \tpas while PA2 and 4 are \cpas.

\begin{table*}[t]
  \centering
  \caption{End-of-term Survey Questions for the Intervention}
  \label{tab:post_survey_questions}
  \footnotesize
  \begin{tabularx}{1\textwidth}{@{} c l X l @{}}
    \toprule
    \textbf{\#} & \textbf{ID} & \textbf{Question} & \textbf{Type} \\
    \midrule
    1 & $Q_{preview\_helpfulness}$ & How helpful was the sneak peek in helping you better understand the scope of the PA? & Likert-scale \\
    \addlinespace[0.3em]
    2 & $Q_{time\_helpfulness}$ & How helpful were the statistics in estimating time spent on PAs? & Likert-scale \\
    \addlinespace[0.3em]
    3 & $Q_{difficulty\_helpfulness}$ & How helpful were the statistics in estimating difficulty rating of PAs? & Likert-scale \\
    \addlinespace[0.3em]
    4 & $Q_{easiness\_TreatPAs}$ & The statistics and sneak peek made it easier for you to follow through with your plans, compared to the PAs that didn’t have them. & Likert-scale \\
    \addlinespace[0.3em]
    5 & $Q_{careful\_planning}$ & The ``Plan'' feature with the preview and statistics helped you think more carefully about how to approach each PA. & Likert-scale \\
    \addlinespace[0.3em]
    6 & $Q_{helpful\_interaction}$ & In what ways did our intervention help you create a better plan, and why? & Open-text \\
    \bottomrule
  \end{tabularx}

  \vspace{0.25em}
  \raggedright\footnotesize
  Note: Q1--Q5 are rated on a 5-point Likert scale (1 = strongly disagree, 5 = strongly agree)
\end{table*}

Data from 252 students were collected through CompassX, including their responses to the planning prompts and self-reported evaluations after completing each PA. 
Additionally, 70 students completed an optional end-of-term survey (Table \ref{tab:post_survey_questions}) to investigate how students perceived the intervention. All analyses were conducted on de-identified data.

\subsection{Evaluation Methods}
\label{sec:data_collection_analysis}

\subsubsection{The Effectiveness of Our Intervention on Planning Behavior and Learning Outcomes} 

We measured the effectiveness of the intervention from two perspectives: 
(1) changes in students' interaction and
(2) associations with learning outcomes.

\textbf{\textit{Changes in Student Interactions.}}
We defined multiple metrics to capture students' interaction changes. The first metric is signed difference (i.e. $\Delta = actual - planned$), on time spent and difficulty rating respectively (i.e. \signedtime and \signeddiffi). They are used to capture directional estimation bias. 

\vspace{-1.5em}

\[
\mathit{\text{Adherence}} =
\begin{cases}
\text{highly-adherent}, & \text{if } \mathrm{avg}(|\Delta time|) < 2 
\\
   & \quad \text{AND } \mathrm{avg}(|\Delta difficulty|) < 1 \\
\text{poorly-adherent}, & \text{otherwise}
\end{cases}
\]
\vspace{-0.9em}

We define plan adherence as the alignment between students’ plans and their execution.
A student is highly-adherent to a group of PAs if, on average, their estimated time spent and difficulty rating on those PAs closely match their actual counterparts. That is, $avg(|\Delta time|)$ and $avg(|\Delta difficulty|)$ are close to 0. 
The thresholds correspond to a one-step deviation in each measurement scale:
time was selected from discrete hourly options (1–12 hours in 2-hour increments), and difficulty was measured on a 5-point Likert scale.


We examined the percentage of highly-adherent students across \cpas and \tpas to determine whether the intervention can assist more students to achieve higher plan adherence.
Lastly, we also compared students’ planned and actual start/end dates for each PA to assess whether they followed their intended timelines. 

\textbf{\textit{Changes in Performance.}}
We compared PA scores and exam scores between highly- (n = 117)  and poorly-adherent (n = 125) students on \cpas only. Students were labeled using \cpas only to capture baseline plan adherence.
We conducted independent-sample t-tests on the scores of PAs, a midterm, and the final grade between the highly-adherent and poorly-adherent group. 

\subsubsection{Students’ Perceptions of the Intervention} 


We investigated the responses to 6 end-of-term survey questions (Table \ref{tab:post_survey_questions}) to understand students’ perceptions of the intervention.
The five Likert-scale questions were examined quantitatively by grouping responses into negative (1–2), neutral (3), and positive (4–5) categories. This allowed us to summarize the overall trends in perceived usefulness. 
Meanwhile, responses to the open-text question were analyzed using inductive thematic analysis \cite{jowsey2021general, naeem2023step}. Two authors independently coded responses based on the 6R coding strategy \cite{naeem2023step}, and iteratively refined codes into broader themes through discussion. 
We also recorded the frequency of each main theme to better understand the most significant factors in students’ feedback.


\section{Impact on Student Interactions and Learning Outcomes}
\label{sec:result_behavior_performance}

In this section, we report quantitative results 
examining how the intervention affected students’ 
behaviors and learning outcomes.

\subsection{More Achievable Planning with the Intervention}
\label{sec:result_5_1_conservative}

\begin{figure}[t]
  \centering
  \includegraphics[width=0.94\columnwidth]{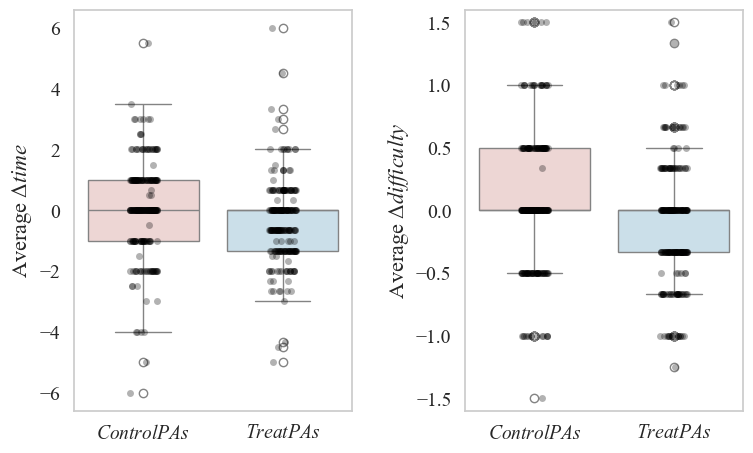}
  \caption{Distribution of signed differences in time estimation and difficulty ratings between \cpas and \tpas.}
  \Description{Distribution of Signed Difference}
  \label{fig:signed_difference}
\end{figure}

The signed differences between \cpas and \tpas showed some statistically significant patterns. As shown in Figure \ref{fig:signed_difference}, students tended to produce more conservative plans for \tpas. \tpas' average \signedtime was lower than \cpas' ($-0.37$ vs. $0.05$ hours), with a smaller standard deviation ($1.31$ vs. $1.49$ hours). Similarly, \tpas' \signeddiffi was lower than \cpas' ($-0.13$ vs. $0.06$) too, with reduced standard deviation ($0.48$ vs. $0.56$). Paired-sample t-tests confirmed these differences are statistically significant as well ($p_{\Delta time}$ = 0.00249, $p_{\Delta difficulty}$ = 0.000011). This indicates that students estimated \tpas' time longer than they actually needed and rated difficulty higher than they actually were. Such conservative planning leaves a larger room for error during the plan execution, making the plans more achievable.

More students were highly-adherent to \tpas as well. 61.54\% of students were labeled highly-adherent to \tpas while only 47.98\% with \cpas.
This 13.56\% increase in high-adherence was also a statistically significant improvement according to a paired t-test ($p = 0.00093$). 
We also conducted a robustness check using a more lenient adherence threshold (\textit{avg($|\Delta time|$) < 3 and avg($|\Delta difficulty|$) < 1}).
The intervention yielded a similar improvement in adherence (11.1\% increase; $p = 0.00346$).

We also found that students’ estimated duration (i.e., the number of days between start and completion) closely matched their actual completion time, indicating improved accuracy in workload estimation. However, no difference was found between \cpas and \tpas with respect to start dates. Regardless of the intervention, students tended to start their PAs later than planned. This suggests that while the intervention improved estimation of task demands, it did not necessarily encourage students to start PAs as planned.


\subsection{Plan Adherence is Associated with Better Learning Outcomes}
\label{sec:5_2_learning_outcome}

\begin{figure}
  \centering
  \includegraphics[width=0.48\textwidth]{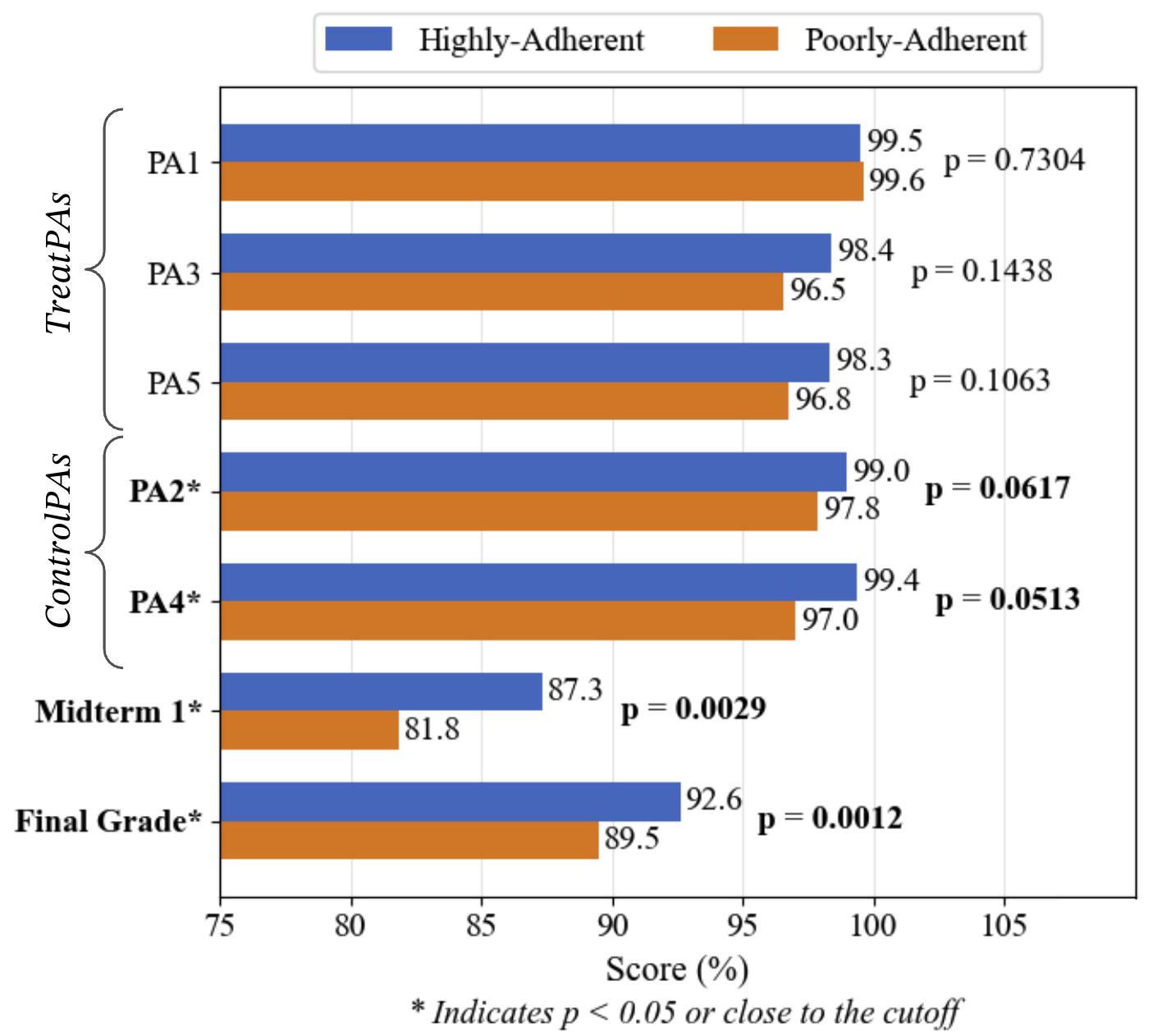}
  \caption{Comparison of assessment scores between Highly- and Poorly-adherent Students}
  \Description{Comparison of Scores between Highly- and Poorly-adherent Students}
  \label{fig:score_by_adherence}
\end{figure}

Figure~\ref{fig:score_by_adherence} visualizes the normalized scores across different course assessments, with respect to highly- and poorly-adherent students based on \cpas. 
There were 117 highly- and 125 poorly-adherent students.
Figure~\ref{fig:score_by_adherence} indicates that the highly-adherent group consistently achieved higher mean scores than the poorly-adherent group on all assessments, but only the gaps in the midterm and in the final grade are statistically significant ( $p$ = 0.0029 and $p$ = 0.0012). This observation might be due to the ceiling effects of PAs, as the class averages were above 95\%. Such high class averages leave limited room to detect differences for analysis. Still, we noticed marginal statistical differences in the \cpas ($p$ = 0.0617 for PA2, $p$ = 0.0513 for PA4). 
These findings suggest that stronger plan adherence was associated with better learning outcomes. 
Given that the intervention increased plan adherence, it may also have the potential to indirectly support better learning outcomes.


\section{Students Perceived the Intervention as Helpful for Planning}
\label{sec:6_result_of_thamatic_analyses}

In this section, we present findings from the end-of-term survey to better understand how students perceived the intervention.

\begin{figure}
  \centering
  \includegraphics[width=0.47\textwidth]{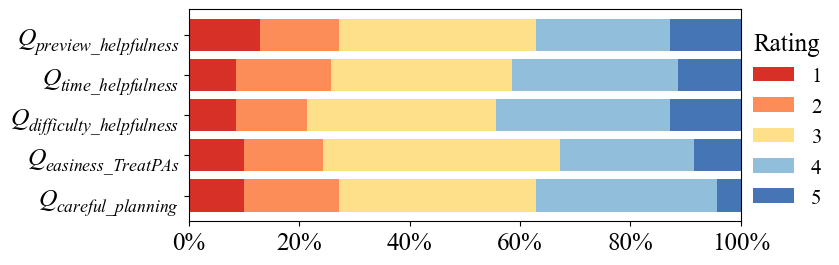}
  \caption{Distribution of Likert-scale Responses from End-of-term Survey. (1 = Strongly Disagree, 5 = Strongly Agree)}
  \Description{Result of Likert-scale Responses from Post-survey}
  \label{fig:rq3_plot}
\end{figure}

Figure~\ref{fig:rq3_plot} summarizes student responses to the Likert-scale questions on the end-of-term survey. Although a good portion of students marked neutral (36.29\% on average), more respondents reported a positive rating (>=4, 38.57\%) than a negative rating (<=2, 25.14\%) across all questions. 
$Q_{difficulty\_helpfulness}$ received the highest percentage of positive responses (44.29\%), indicating that students found the difficulty estimation interaction the most helpful. 
In contrast, $Q_{easiness\_TreatPAs}$ received the largest proportion of neutral-to-negative responses (67.14\%), suggesting uncertainty about the intervention's ability to support plan execution, consistent with students starting PAs later than planned in both \tpas and \cpas.

\begin{table*}[htbp]
\caption{Thematic Analysis Results for $Q_{helpful\_interaction}$. Four themes and their frequencies are reported, along with definitions and sub-themes. Each sub-theme is also characterized by its nuance, positive (\textcolor{green}{+}) vs. negative (\textcolor{red}{-}).}
\label{tab:thematic_analysis_Q6}
\footnotesize

\centering
\begin{tabular}{p{4cm} p{7cm} p{5.5cm}}
\toprule
\textbf{Theme} (Frequency) & \textbf{Definition} & \textbf{Sub-theme}\\
\midrule
\textbf{Practical Planning Support} (46) & The intervention provided students with structured guidance to create executable plans by supporting accurate time estimation, anticipating task difficulty and scope, and aligning workload expectations and forming executable plans. & 
\begin{tabular}[t]{@{}p{5.5cm}@{}}
    - Time estimation and management support (\textcolor{green}{+})\\
    - Difficulty anticipation and scope understanding support (\textcolor{green}{+})\\
    - Workload expectation support (\textcolor{green}{+})
\end{tabular} \\

\midrule

\textbf{Mental Benefit} (4) & Emotional or psychological effects students experienced as a benefit. & 
\begin{tabular}[t]{@{}l@{}}
    - Mental help (\textcolor{green}{+})
\end{tabular} \\
\midrule

\textbf{Perceived Applicability Issue} (27) & Factors that students think limit the applicability of the intervention. & 
\begin{tabular}[t]{@{}p{5.5cm}@{}}
    - Individual experience and differences (\textcolor{red}{-})\\
    - Personal planning style \& preferences (\textcolor{red}{-})\\
    - Information not perceived as useful (\textcolor{red}{-})
\end{tabular} \\

\midrule

\textbf{Adoption Challenge} (9) & Behavioral or motivational factors prevent students from engaging with or effectively applying the intervention. & 
\begin{tabular}[t]{@{}p{5.5cm}@{}}
    - Behavioral or situational barriers (\textcolor{red}{-})\\
    - Disengagement / Cognitive disengagement (\textcolor{red}{-})
\end{tabular} \\

\bottomrule
\end{tabular}

\vspace{0.25em}
  \raggedright\footnotesize
  (\textcolor{green}{+}) : positive, (\textcolor{red}{-}) : negative

\end{table*}

Our thematic analysis results add more insights to the Likert-scale question responses about what aspects students found the intervention helpful. 
Table~\ref{tab:thematic_analysis_Q6} lists the four main themes and 9 sub-themes identified from responses to $Q_{helpful\_interaction}$. 
Consistent with the Likert-scale results, 46 students mentioned the intervention provided practical planning support, such as improving their ability to estimate time requirements and anticipate difficulty level and scope of work. 
\begin{quote}
     "\textit{Yes, I think these supports helped me create a more realistic plan because they gave me an idea of what to expect when I do the PAs myself, such as how difficult they're going to be or how much time I'll spend on them, so I can create a good plan for myself.}"
\end{quote}

Even mental benefits were mentioned by several students as additional effect of our intervention. Their responses indicated that seeing data from past students reduced feeling guilty about time spent and helped validate their own experience. Others mentioned that the intervention helped them build confidence in their planning mindset, making them feel less anxious about underestimating.

\begin{quote}
    "\textit{I feel they were more helpful in determining a mental plan and mindset.}"
\end{quote}
\begin{quote}
    "\textit{Yes, seeing how other students felt about the PA both validated my experience and helped me understand the time and difficulty of it.}"
\end{quote}

However, some students noted that the intervention did not necessarily help them generate more realistic plans. Their feedback highlighted concerns about the relevance of the statistics, as cohort-level data did not always reflect their individual circumstances, learning styles, or planning strategies. 
In addition, 9 students described behavioral or situational challenges (e.g., procrastination, busy schedules) that limited their ability to adopt the intervention. These perspectives align with the mixed responses to the final two survey questions related to plan execution in Figure~\ref{fig:rq3_plot}, where some students were skeptical about the intervention's effectiveness.

\begin{quote}
    "\textit{Potentially, but I think what most helped me inform my plan was my past experiences with the PAs...}"
\end{quote}
\begin{quote}
    "\textit{Not really, I like to go at my own space and did most of the PA's faster than other people said it took them}"
\end{quote}
\begin{quote}
    "\textit{The statistic wasn't as helpful for me simply because I had a busy schedule this quarter and tried to work on PAs whenever I had the time.}"
\end{quote}

\section{Discussions}

Our findings suggest that historical cohort data can support more achievable planning in computing courses. 
Below, we discuss three implications of the data-driven planning support.


\emph{Impact of the Intervention.}
Despite its lightweight design, our intervention led to observable changes in students' task estimation. 
Specifically, students produced more conservative estimates and generated more achievable plans (Section~\ref{sec:result_5_1_conservative}). 
These behavioral changes suggest that providing any specific expectations on assignments can recalibrate students' estimation about programming assignment demands in an authentic classroom setting.
Considering how influential historical statistics could be, presenting slightly conservative past statistics may be beneficial in computing courses, especially given prior findings that younger generations \cite{dunning2004flawed} and poorly performing students \cite{falchikov1989student, pieterse2017students} often overestimate their abilities. 

In addition, the intervention brings mental benefits by shaping students' affective experience. 
Clearer expectations and visibility into peers' past efforts reassured students, validated their experiences, reduced feelings of isolation, and supported more accurate self-evaluation for planning, consistent with prior literature on social comparison \cite{festinger1954theory, dijkstra2008social}.


\emph{Building Trust in Data-Driven Intervention.}
A key challenge identified from the thematic analysis (Section~\ref{sec:6_result_of_thamatic_analyses}) is that not all students trusted or found the cohort-level statistics applicable to their own situations. 
While individual differences certainly exist, the statistics were derived from a large dataset of students from the same course, making it likely that comparable peers were represented. 
Nevertheless, some students did not perceive the data that way, which appeared to limit their engagement with the intervention.

This finding suggests that transparency is essential for maintaining trust and engagement in interventions. 
Future systems should clearly communicate how the data were collected and emphasize that the statistics reflect a diverse range of student experiences. 
Moreover, combining cohort-level benchmarks with students' own historical data may support more personalized decision-making.

\emph{The Planning--Execution Gap.}
While our intervention helped students create more achievable plans, executing those plans remained challenging. 
We observed that many students still started assignments later than planned (Section~\ref{sec:result_5_1_conservative}).
This suggests that creating an achievable plan is only the first step; successfully executing that plan may require additional support.
This gap may also help explain why plan adherence was associated with better learning outcomes. Students who achieved higher adherence may have benefited not only from creating more realistic plans, but also from their ability to follow through with them. 
Future work should explore how planning support can be combined with interventions that encourage plan execution and sustained engagement.

\section{Threats to Validity}

One limitation of this study is that the intervention's observed effects may have been confounded by some potential factors, such as differences in assignment complexity or cohort characteristics.
While the intervention was associated with changes in students' planning behaviors and learning outcomes, these effects were correlational rather than causal, as other factors may have contributed to these observations.
Another limitation is that our intervention’s historical data was solely based on a single cohort’s statistics, and the study was conducted in a single computing course, which may limit generalizability. 
In addition, the measure of actual time spent and difficulty ratings were self-reported by students, which can be subjective. 
Students may also have adjusted their responses toward perceived normative values after viewing the historical statistics.
These factors could have affected the accuracy of our adherence metric and results. 
Additionally, the optional end-of-term survey had a relatively low response rate (70/252), potentially introducing response bias and limiting the representativeness.

We encourage future studies in other courses to explore integrating data-driven planning interventions into their workflows. This will help validate the robustness, scalability, and broader applicability of the data-driven planning support.





 \section{Conclusion}

In this work, we designed and deployed a low-cost, data-driven intervention that integrates assignment previews and cohort-level historical statistics into the planning phase of the programming assignment workflow. In a real-world course deployment with 252 students, the intervention helped produce more conservative and achievable plans. Higher adherence was also associated with stronger learning outcomes. End-of-term survey responses indicated that most students found the intervention helpful, emphasizing both its practical planning support and its mental benefits.

Our findings suggest that such task-specific historical data benefits students as they clarify expectations on each task like an assignment. They also highlight the importance of supporting the transition from planning to execution and building trust in data-driven systems through transparent design. While our study was conducted in a computing course, the core idea of embedding historical benchmarks into students' learning workflows may extend to other larger-scale educational contexts.



\bibliographystyle{ACM-Reference-Format}
\bibliography{citation}

\end{document}